\documentstyle[11pt,]{article}
\begin{document}
{\bf{SPECTRA OF RELIC GRAVITONS AND BRANS-DICKE THEORY}}
\vspace{0.2in}\\
\centerline {B.K.Sahoo $^{1}$} 
\centerline{Department of Physics,}
\centerline{Utkal University, Bhubaneswar-751004, India}
\centerline{$^{1}$ bijaya@iopb.res.in}
\vspace{0.25in}
\begin{abstract}
The spectra of relic gravitational waves produced as a result of
cosmological expansion of the inflationary models are derived in
Brans-Dicke (BD)theory of gravity.The time dependence of the very early 
Hubble parameter and matter energy density are derived from  
frequency dependent spectrum of relic gravitational waves.Also it is
found that Brans-Dicke scalar field contributes to the energy 
density of relic gravitons.
\end{abstract}
PACS NO: 98.80.-k,98.80.cq
\section{INTRODUCTION}
Relic gravitational waves are, quite possibly, the only source of
information on the physical conditions in the very early universe.The energy 
density and spectrum of relic gravitational waves depend on a specific rate of
 expansion of the universe in its very distant past.The actual detection of 
relic gravitons, or even the experimental restriction of their possible spect
ral energy density, are capable of producing fairly meaningfull
conclusions about the parameters of the very early universe[1].In general,
the
information 
stored in the relic gravitons allows one to judge the overall
evolution of 
the cosmological scale factor.However,in more specific models of the
universe,
this 
information gives us the direct data on the variability of the Hubble
parameter.From the theoretical point of view, the most natural
alternatives to General relativity(GR) are the scalar-tensor
theories,which contain an additional scalar field,$\phi$,whose
relative importance is determined by the arbitrary coupling function
$\omega(\phi)$ [2].GR corresponds to the case where the field
$\phi$=const.=1 implying an infinite $\omega(\phi)$.The Brans-Dicke
theory is defined by a constant $\omega(\phi)$[3].This is a successful 
theory because it explains almost all the important features of the
evolution of the universe.Some of the problems like inflation,early
and late time behaviour of universe[4], cosmic acceleration and
structure formation[5], cosmic acceleration,and
coincidence problem[6] can be explained in the B-D formalism.For large 
$\omega$,B-D theory gives the correct amount of inflation and early
and late time behaviour,and for small negative $\omega$ it correctly
explains cosmic acceleration,structure formation and coincidence
problem.Also,this theory appears naturally in supergravity
theory, Kaluza-Klein theories and in all the known effective string
actions and thus,in itself, justifies further consideration.

The fundamental feature of BD theory is that there is a scalar field
 $\phi$ coupled to the curvature.Its essential role in homogeneous
 cosmologies is to induce a time variation of the gravitational
 constant.Also this scalar field $\phi$ modifies the expansion rate in 
 the early universe.Cosequently, flatness problem,horizon problem, the
 primodial nucleosynthesis problem,and the energy density and spectrum 
 of relic gravitational waves  are all modified.Here we aim to assess
 the modifications induced by the coupling of the scalar field to the
 energy density and spectrum of the graviton.
  Previous work on the subject of gravitational radiation within
  Brans-Dicke theory, carried out by Wagoner[7] was concerned with
  gravitational waves generated in the weak field
  limit.Later, Barrow[8] derived the gravitational wave equation for
  the general class of scalar tensor theories and solved for vacuum
  and radiation era  for different cosmological models.Also Sahoo and
  Singh[9] solved the same wave equation for all epochs of the
  universe and found that  for negative present values of the parameter $\omega$
  gravitational waves grow in all epochs of the universe.\\

     An outline of the paper is as follows.In the next section we
     write the solutions of the gravitational wave equation in BD
     theory .The Bogoliubov transformation coefficients 
     and energy density of gravitons are derived in the section
     three.In the section four frquency dependence of early Hubble
     parameter and matter energy density are derived.We conclude with
     section five.
\section{BASIC EQUATIONS}
The perturbations $h_{ij}$ to the metric that represent weak
gravitational waves can be expressed as
\begin{equation}
h(t,x)=\int d^{3}n h_{n}(t,x)
\end{equation}
\begin{equation}
h_{n}(t,x)=\frac{1}{a^{2}(t)} Y_{n}(t) U_{n}(x)
\end{equation}
The wave number ,$n$,$
n^{2}=(n_{1})^{2}+(n_{2})^{2}+(n_{3})^{2}$,is associated with the
frequency $\nu $(measured in Hz) according to the relation $\nu=
\frac{n}{2\pi a } $ in  units $c=G=\hbar=1$.The functions $ U_{n} $ and
$Y_{n}$ satisfy 
\begin{equation}
(\nabla^{2}+n^{2})U_{n}=0
\end{equation}
\begin{equation}
\ddot{Y_{n}} +\left[ \frac{\dot{\phi}}{\phi}-\frac{\dot{a}}{a} \right] 
\dot{Y_{n}} + \left[ \frac{n^{2}}{a^{2}}-2\frac { \ddot{a}}{a}
  -2\frac{\dot{a}}{a}
  \frac{\dot{\phi}}{\phi}-\frac{d\omega/d\phi}{2\omega+3} \frac
  {\dot{\phi}^{2}}{\phi} + \frac{ \phi dU/d\phi +
    2(\omega+1)U}{2\omega+3} \right]Y_{n}=0
\end{equation}
In the above equations $a(t)$ is the scale factor,$\nabla ^{2}$ is the 
spatial Laplacian operator, an overdot indictes derivative  with
respect to the synchronous cosmic time t.Equation(4) is gravitational
wave equation for zero curvature FRW Universe[8].The Brans-Dicke
scalar-tensor theory of gravity is characterised by the restriction
that $\omega(\phi)=\omega=$ constant and $ U= dU/d\phi=0$.We introduce the conformal time
$\eta$ defind by 
\begin{equation} a(\eta) d\eta = dt.
\end{equation}
Now each component $h_{n}$ of the gravitational wave perturbations can 
be written as 
\begin{equation}
h_{n}(\eta,x)= \frac{1}{R(\eta)} \mu_{n}(\eta) U_{n}(x),
\end{equation}
where,
\begin{equation}R(\eta)=a(\eta) \phi ^{1/2}(\eta).
\end{equation}
So,for the Brans-Dicke scalar-tensor theory of gravity and for
Eq(6), the gravitational wave equation(4) reduces to
\begin{equation}
\mu_{n}^{\prime \prime}(\eta) + \left[ n^{2} - V(\eta) \right]
\mu_{n}(\eta)=0,
\end{equation}
where $V(\eta)= \frac{R^{\prime \prime}}{R}$ and prime denotes the
derivative with respect to $\eta$.Equation(8) looks like the
schrodinger equation for a particle having the potential energy
$V(\eta)$.For high frquency waves or for waves such that $ n^{2}>>
\frac{R^{\prime \prime}}{R}$ the general solution to Eq(8) has the
form
\begin{equation}
\mu_{n}= \frac{1}{\sqrt{2}} \left( \alpha_{n} e^{-in\eta}  +
  \beta_{n}  e^{in\eta}  \right),
\end{equation}
Where $\alpha_{n}$ and $ \beta_{n}$ are arbitrary complex numbers. In
the opposite limit, $ n^{2} << \frac{R ^{\prime \prime }}{R}, $the
general solution is 
\begin{equation}
\mu_{n}= A_{n} R  + B_{n}R \int_{\eta_{i}}^{\eta_{f}} \frac{d\eta}{R}
\end{equation}
Here also $A_{n}$ and $B_{n}$ are arbitrary complex numbers.
\section{GRAVITON SPECTRA }
It is known that a travelling  gravitational wave passing through the
barrier $ V(\eta)$ will always be amplified[10].In the quantum
treatment of the problem,one says that the initial vacuum state of
gravitons goes over to a final multi-particle quantum state - the
particle creation takes place.The final quantum state belongs to a
class of the so-called squeezed quantum states[11].
It is clear from Eq(8) that the behavior of $\mu_{n}$ is determined by 
  the  potential $V(\eta)=\frac{R^{\prime \prime}}{R}$.The solution(9) 
  is valid outside the potential barrier $ V(\eta)$ while the solution 
  (10) applies to the region where the function $V(\eta)$ dominates
  over $n^{2}$.The values of $\eta$ where the regimes (9) and (10)
  interchange are determined by the condition
\begin{equation}
n^{2}=|V(\eta)|
\end{equation}
A wave with a given ``$n$'' enters the barrier region and leaves it in
the ``turning points'' defined by eq(11).We will denote these points
by $\eta_{i}$  and $\eta_{f}$.The indices  ``$i$'' and  ``$f$'' will also
be used 
to 
distinguish the values of the various functions at the initial and
final turning
points. We will match solutions (9) and (10) at $\eta_{i}$ and
$\eta_{f}$ for continuity [12,13]. To the left of the barrier,that is for $ \eta
< \eta_{i}$,  the solution is taken in the form 
\begin{equation}
\mu_{n}=(2n)^{-1/2} e^{in\eta}
\end{equation}
To the right of the barrier,that is for $ \eta > \eta_{f},$ the
solution  has the form(9) where the numerical  values of the co-efficients
 $ \alpha_{n} $ and  $ \beta_{n} $  follow from the joining 
conditions.The actual values of  $ \alpha_{n} $ and $ \beta_{n} $ are
\begin{equation}
\alpha_{n}=\frac{ e^{in(\eta_{i}+\eta_{f})}}{2in}
  \left[ \frac{R_{i}^{\prime}}{R_{f}} - \frac{R_{f}^{\prime}}{R_{i}} + 
    in \left( \frac{R_{f}}{R_{i}}-\frac{R_{i}}{R_{f}}\right) \right]
\end{equation}

\begin{equation}
\beta_{n}=\frac{e^{-in(\eta_{i}-\eta_{f})}}{2in}  \left[
  -\frac{R_{i}^{\prime}}{R_{f}} + \frac{R_{f}^{\prime}}{R_{i}}
+ in\left( \frac{R_{f}}{R_{i}} + \frac{R_{i}}{R_{f}} \right) 
\right]
\end{equation}
As $R(\eta)= a(\eta) \phi^{1/2}(\eta)$ ,one has
\begin{equation}
 R^{\prime}=\phi^{1/2} H a^{2} + \frac{\dot{\phi} a^{2}}{2\phi^{1/2}}
\end{equation}
where $H=\frac{\dot{a}}{a}$ is used.We write below the complete expression for
$\alpha_{n}$ and $\beta_{n}$.
\begin{equation}
\alpha_{n}= \frac{ e^{in(\eta_{i}+\eta_{f})}}{2in} \left[ \frac{ H_{i} 
    a_{i}^{2}}{a_{f}} \frac{\phi_{i}^{1/2}}{\phi_{f}^{1/2}} +
  \frac{a_{i}^{2}}{a_{f}} \frac{\dot{\phi_{i}}}{2\phi_{f}^
{1/2}\phi_{i}^{1/2}} -
\frac{H_{f}a_{f}^{2}}{a_{i}}\frac{\phi_{f}^{1/2}}{\phi_{i}^{1/2}} -
\frac{a_{f}^{2}}{a_{i}} \frac{\dot{\phi_{f}}}{2 \phi_{i}^{1/2}
  \phi_{f}^{1/2}} +in \left( \frac{a_{f}}{a_{i}}
  \frac{\phi_{f}^{1/2}}{\phi_{i}^{1/2}} -
  \frac{a_{i}}{a_{f}}\frac{\phi_{i}^{1/2}}{\phi_{f}^{1/2}}\right)
\right]
\end{equation}

\begin{equation}
\beta_{n}= \frac{ e^{in(\eta_{i}-\eta_{f})}}{2in} \left[ -\frac{ H_{i} 
    a_{i}^{2}}{a_{f}} \frac{\phi_{i}^{1/2}}{\phi_{f}^{1/2}} -
  \frac{a_{i}^{2}}{a_{f}} \frac{\dot{\phi_{i}}}{2\phi_{f}^
{1/2}\phi_{i}^{1/2}} +
\frac{H_{f}a_{f}^{2}}{a_{i}}\frac{\phi_{f}^{1/2}}{\phi_{i}^{1/2}} +
\frac{a_{f}^{2}}{a_{i}} \frac{\dot{\phi_{f}}}{2 \phi_{i}^{1/2}
  \phi_{f}^{1/2}} +in \left( \frac{a_{f}}{a_{i}}
  \frac{\phi_{f}^{1/2}}{\phi_{i}^{1/2}} +
  \frac{a_{i}}{a_{f}}\frac{\phi_{i}^{1/2}}{\phi_{f}^{1/2}}\right)
\right]
\end{equation}
In fact,we have calculated the coefficients of the Bogoliubov
transformation which relates the in and out quantum states of
gravitons[14]. Let us recall that in the quantum theory the quantity
$|\beta_{n}|^{2}$ defines the spectral energy density of gravitons
created from the in vacuum state.The contepmorary energy density of
gravitons integrated over some frequency interval is given by the
formula[13,14,15]
\begin{equation}
 E= \int E(\nu) d\nu 
\end{equation}
The density of states is $dN=\frac{ 4\pi\nu^{2}}{c^{3}} d\nu $ and the 
corresponding $ E(\nu)d\nu $ summed over two polarization states is
given by \\
\begin{equation}
E(\nu) d\nu = 8\pi \nu^{3} d\nu |\beta_{n}|^{2},
\end{equation}
with $\hbar$=c=1.
The remarkable property of Eq(16) and Eq(17) is that they contain only 
the initial ``$i$'' and final ``$f$'' values of  a, H and  $\phi$.Equations(16)
and (17) can be further simplified for particular numerical values of 
$H_{i}a_{i}$, $H_{f}a_{f}$  and  $\frac{a_{f}}{a_{i}}$.For instance,in the 
case of $(H_{i}a_{i})^{2}) >> n^{2} $  and  $(H_{f}a_{f})^{2} >>
n^{2}$,one obtains[12]

\begin{equation}
\alpha_{n}= \frac{ e^{in(\eta_{i}+\eta_{f})}}{2in} \left[ \frac{ H_{i} 
    a_{i}^{2}}{a_{f}} \frac{\phi_{i}^{1/2}}{\phi_{f}^{1/2}} +
  \frac{a_{i}^{2}}{a_{f}} \frac{\dot{\phi_{i}}}{2\phi_{f}^
{1/2}\phi_{i}^{1/2}} -
\frac{H_{f}a_{f}^{2}}{a_{i}}\frac{\phi_{f}^{1/2}}{\phi_{i}^{1/2}} -
\frac{a_{f}^{2}}{a_{i}} \frac{\dot{\phi_{f}}}{2 \phi_{i}^{1/2}
  \phi_{f}^{1/2}} 
\right]
\end{equation}

\begin{equation}
\beta_{n}= \frac{ e^{in(\eta_{i}-\eta_{f})}}{2in} \left[ -\frac{ H_{i} 
    a_{i}^{2}}{a_{f}} \frac{\phi_{i}^{1/2}}{\phi_{f}^{1/2}} -
  \frac{a_{i}^{2}}{a_{f}} \frac{\dot{\phi_{i}}}{2\phi_{f}^
{1/2}\phi_{i}^{1/2}} +
\frac{H_{f}a_{f}^{2}}{a_{i}}\frac{\phi_{f}^{1/2}}{\phi_{i}^{1/2}} +
\frac{a_{f}^{2}}{a_{i}} \frac{\dot{\phi_{f}}}{2 \phi_{i}^{1/2}
  \phi_{f}^{1/2}}
\right]
\end{equation}
On the otherhand,if $(H_{i}a_{i})^{2}\approx n^{2}, (H_{f}a_{f})^{2}
\approx n^{2}$,and $\frac{a_{f}}{a_{i}} >> 1$ one  derives

\begin{equation}
\alpha_{n}= \frac{ e^{in(\eta_{i}+\eta_{f})}}{2in} \left[-
\frac{H_{f}a_{f}^{2}}{a_{i}}\frac{\phi_{f}^{1/2}}{\phi_{i}^{1/2}} -
\frac{a_{f}^{2}}{a_{i}} \frac{\dot{\phi_{f}}}{2 \phi_{i}^{1/2}
  \phi_{f}^{1/2}} +in \frac{a_{f}}{a_{i}}
  \frac{\phi_{f}^{1/2}}{\phi_{i}^{1/2}}
\right]
\end{equation}

\begin{equation}
\beta_{n}= \frac{ e^{in(\eta_{i}-\eta_{f})}}{2in} \left[ +
\frac{H_{f}a_{f}^{2}}{a_{i}}\frac{\phi_{f}^{1/2}}{\phi_{i}^{1/2}} +
\frac{a_{f}^{2}}{a_{i}} \frac{\dot{\phi_{f}}}{2 \phi_{i}^{1/2}
  \phi_{f}^{1/2}} +in  \frac{a_{f}}{a_{i}}
  \frac{\phi_{f}^{1/2}}{\phi_{i}^{1/2}}
\right]
\end{equation}
We will consider in more detail the case where the inflationary stage
ends with a rapid transition to the radiation-dominated stage.The
relevant barrier in Eq(8) is described by the function $V(\eta)$ which 
first increases with $\eta$ and then decreases sharply up to zero.Let
us consider the waves interacting with this barrier.Their wave numbers 
obey the conditions $(H_{i}a_{i})^{2} \approx n^{2},(H_{f}a_{f})^{2}>> 
n^{2} $ and we assume also  $\frac{a_{f}}{a_{i}}>> 1$.For acceptable
inflationary models these waves have a present day frequencies in the
interval $ 10^{8} Hz < \nu  <  10^{-16} Hz$ [12].From eq(16) and
eq(17) we obtain
\begin{equation}
\alpha_{n}= \frac{ e^{in(\eta_{i}+\eta_{f})}}{2in} \left[-
\frac{H_{f}a_{f}^{2}}{a_{i}}\frac{\phi_{f}^{1/2}}{\phi_{i}^{1/2}} -
\frac{a_{f}^{2}}{a_{i}} \frac{\dot{\phi_{f}}}{2 \phi_{i}^{1/2}
  \phi_{f}^{1/2}}
\right]
\end{equation}

\begin{equation}
\beta_{n}= \frac{ e^{in(\eta_{i}-\eta_{f})}}{2in} \left[ +
\frac{H_{f}a_{f}^{2}}{a_{i}}\frac{\phi_{f}^{1/2}}{\phi_{i}^{1/2}} +
\frac{a_{f}^{2}}{a_{i}} \frac{\dot{\phi_{f}}}{2 \phi_{i}^{1/2}
  \phi_{f}^{1/2}} 
\right]
\end{equation}
And 
\begin{equation}
|\alpha_{n}|^{2}=|\beta_{n}|^{2}=\frac{1}{4n^{2}}H_{i}^{2}H_{f}^{2}a_{f}^{4}
\left[ \frac{\phi_{f}}{\phi_{i}} + \frac{1}{\phi_{i}}( \frac{\dot{\phi_{f}}}{H_{f}}) + \frac{1}{4\phi_{i} \phi_{f}}(\frac{\dot{\phi_{f}}}{H_{f}})^{2}\right]
\end{equation}
Using $|\beta_{n}|^{2}$ in eq(19) with $ \nu=\frac{n}{2\pi a},$ we derive\\
\begin{equation}
E(\nu) d\nu = \frac{1}{(2\pi)^{3}} \frac{1}{a^{4}} H_{i}^{2} H_{f}^{2} 
a_{f}^{4} \left[ \frac{\phi_{f}}{\phi_{i}} + \frac{1}{\phi_{i}}( \frac{\dot{
\phi_{f}}}{H_{f}}) + \frac{1}{4\phi_{i} \phi_{f}}(\frac{\dot{\phi_{f}}}{H_{f}}
)^{2}\right]
\frac{d\nu}{\nu}
\end{equation}

The difference between this expresion and the expression in ref[12
] is the presence of contributions due to the scalar field .The 
kinetic energy of the BD scalar field makes a non-zero contribution to 
$E(\nu)$ and this comes from the rapid variation of $\phi$ during the
inflation epoch.When $\phi $=constant=1, this expression reducess to the 
expression obtained by Grishchuk and Solokhin[12].The extra terms are
due to the rapid variation of $\phi$ during the inflationary era.
\section{HUBBLE PARAMETER AND MATTER ENERGY DENSITY}
The entire dependence on $n$(and
hence ,on frequency $\nu$) is contained in the factor $H_{i}^{2}$.In its
turn,$H_{i}^{2}(n)$ is proportional to the matter energy density at
the inflationary stage , namely, at those instants of time $t_{n}$,when
the waves with the corressponding wave numbers ``$n$'' were entering the
under barrier region   $H_{i}^{2}(n)= \frac{8\pi}{3} \rho(t_{n}) $. The value of 
$H_{f}^{2}$ is propertional to the matter energy density at the end of 
inflation.Some part $\delta$ of this energy($\delta\approx1$) has been 
transformed into the electromagnetic radiation.In the course of
cosmological expansion the energy density of the radiation was
decreasing in proportion to $(\frac{a_{f}}{a})^{4}$.The present day
energy density $E_{\gamma}$ of the radiation ($3^{0}$K microwave background)
is related to $H_{f}^{2}$ according to
\begin{equation}
H_{f}^{2}= \frac{8\pi}{3} \rho(t_{f})=\frac{8\pi}{3}.\frac{1}{\delta}( 
\frac{a}{a_{f}})^{4}E_{\gamma}
\end{equation}
Using the expression for $H_{i}^{2}$ and $H_{f}^{2}$,neglecting the numerical factors of order  unity, and restoring $\rho_{p}$ for the correct dimensionality, we obtain finally \\
\begin{equation}
E(\nu)=E_{\gamma} \frac{\rho(t_{n})}{\rho_{p}}
\left[ \frac{\phi_{f}}{\phi_{i}} + \frac{1}{\phi_{i}}( \frac{\dot{
\phi_{f}}}{H_{f}}) + \frac{1}{4\phi_{i} \phi_{f}}(\frac{\dot{\phi_{f}}}{H_{f}}
)^{2}\right]
\end{equation}
This very simple expression allows us to link the energy density $E(\nu)$ of relic gravitational waves measurable today with the time dependent values of the matter density ``$\rho$'' and the Hubble parameter ``H'' attributed to the very early universe .One can say, a little loosely,that the function $E(\nu)$ stores the information on the rate of expansion of the universe in that distant past when the relic waves with frequencies $\nu$ first started to emerge from zero-point quantum fluctuations.
If the expansion law $a(t)$ were precisely known, Eq.(27) would give us a definite spectral dependence of $E(\nu)$.For example, for the power-law scale factors $a(t) \sim t^{\alpha}$, one has $ H(t)=\alpha t^{-1}$.From the condition $H_{i}(t_{n}) a_{i}(t_{n})=n$ we derive  $ n \sim t_{n}^{\alpha - 1}$, that is ,  $t_{n} \sim n^{ \frac{1}{\alpha -1}}$. Since $ \rho(t_{n}) \sim H_{i}^{2}(t_{n}) \sim t_{n}^{-2} \sim n^{\frac{-2}{\alpha -1}}$ one can obtain $E(\nu) \sim \nu ^{\frac{-2}{\alpha-1}} $, that is the power law frequency dependence for $E(\nu)$.In the case of strictly de Sitter expansion($H_{i}=$constant) one would obtain the spectrum $E(\nu) = $ constant.Equation (29) gives $\rho$ and  H  as functions of frequency $\nu $,\\
\begin{equation}
\rho(\nu)= \frac{\rho_{p}}{E_{\gamma}} \frac{E(\nu)}{\phi_{c}},
\end{equation}

\begin{equation}
H^{2}(\nu) = \frac{8\pi}{3} \rho(\nu),
\end{equation}

where $\phi_{c}$ is contribution due to the presence of scalar field
$\phi$[square bracket term in eq(29)]and,therefore,  gives $\rho$ and H as nonmanifest functions of time,since every $\nu $ corresponds to some $t_{n}$.

\section{CONCLUSION}
We have derived the energy density and spectrum of gravitons in the
context of Brans-Dicke theory.It is found that due to the presence of
scalar field $\phi$ , the expression for graviton energy
density is modified[12].The kinetic energy of $\phi$ enters into the
expression .This happens due to rapid variation of the scalar field
$\phi$ during the inflationary era .
Also the time dependence of the early Hubble parameter and
matter energy density are derived we find  that the frequency
dependence of the relic graviton energy density spectrum replicates
exactly the time dependence of the very early Hubble parameter.  
\centerline{{\bf Acknowledgement}}
The authors thank Institute of Physics,Bhubaneswar,India for providing facility of the computer center.

\end{document}